\documentstyle{article}
\newcommand{\be}{\begin{equation}}
\newcommand{\ee}{\end{equation}}

\newcommand{\la}[1]{\label{#1}}
\newcommand{\r}[1]{(\ref{#1})}
\renewcommand{\c}[1]{\cite{#1}}
\begin{document}
\title{Semileptonic decays of heavy to light mesons from\\
an MIT bag model.\thanks{Work supported by the KBN
grant no.\ 20-38-09101.}
}
\author{
{\sc M.\,Sadzikowski}\thanks{E-mail {\sc ufsadzik@ztc386a.if.uj.edu.pl}}\\
{\it Institute of Physics, Jagellonian University,} \\
{\it ul.\,Reymonta 4, PL-30\,059 Krak\'ow, Poland.}}
\date{}

\maketitle
\begin{center}

\begin{abstract}

Using the (modified) MIT bag model we calculated formfactors and decay
widths for pseudoscalar - pseudoscalar and pseudoscalar - vector semileptonic
decays of heavy to light mesons. We discuss the physical
phenomena which are important in these processes and their influence on the
measurable quantities. Our results are consistent with the available
experimental
data. A comparison with the results of other models is also given.

\end{abstract}
\end{center}
\newpage
\section{Introduction}
There is a wide variety of interesting questions concerning semileptonic decays
of heavy mesons into light ones. There was a lot of papers
dealing with this subject and trying to resolve problems from different
points of view (for example  \c{wir, bau, korn, isg, lub1, bern, lub2, cas1,
cas2, cas3, ball, aba, hai, du, nar}), because the first principle approach
fails for the obvious reason that nonperturbative effects of strong
interactions
affect the picture of the weak decay. Even the Heavy Quark Symmetry (HQS), such
a powerful method in many other situations, can only relate the hadronic matrix
elements of $D$ decays to the corresponding ones of $B$ decays \c{hqs}. For
this
reason pictures consistently describing possible semileptonic
processes (within models) are still of importance.

An example of an interesting problem is to try to understand
the suppression of the ratios of the decay widths between
pseudoscalar --- vector and pseudoscalar --- pseudoscalar transitions.
For example:
\be
\la{R}
R^{DK^{\ast }}_{DK} \equiv \frac{\Gamma (D \rightarrow K^{\ast } e \nu_e)}{
\Gamma (D \rightarrow K e \nu_{e})}
\ee
is about 0.5
although in the case of $R^{BD^\ast }_{BD}$, when a heavy meson B decays into
a D meson, experiments give us a number about six time larger. This fact is
clearly
connected with some nonperturbative phenomena in strong interactions. Another
interesting question concerns the relation between such processes as
$D^0 \rightarrow K^{\ast -}e^+\nu_{e}$ and $D_s \rightarrow \phi e^+\nu_{e}$.
{}From the experimental results \c{pdg}\footnote{see Table III}, taking the
center
value and forgetting about the large errors, it seems that both processes are
very close in probability in spite of the fact that they differ by the
spectator antiquark. This is in agreement with the naive $SU(3)$ flavour
symmetry expectations, but we know that this symmetry is broken.
Another unsolved problem is connected with the $q^2$ - dependence of the
formfactors. Some models assume the pole dependence of the formfactors
(which we also do), others try to check this dependence (for example \c{lub1,
lub2, ball, nar}). It is generally believed that in $D$ decays the pole form
dependence is quite reasonable, but in the case of $B$ decays the question is
still open \c{ball}, \c{nar}.

Finally, the values of the decay widths also are useful for
the determination of the CKM matrix element $|V_{ub}|$ as well as are probes
of our understanding of semileptonic processes.

In this paper we propose to apply the simplest version of the corrected MIT bag
model (\c{shur}, \c{iza}).
This model was successfully applied to a calculation of the Isgur-Wise function
\c{sadzal} and now proves also promising in the case of the heavy to light
transitions. The results agree well with the available data and the model also
gives some predictions for other not yet measured processes.

\section{Formfactors}

In the case of pseudoscalar - pseudoscalar transitions the matrix elements of
the hadronic currents can be described in terms of two invariant formfactors:
\be
\la{pp}
\langle P(p')|V_{\mu}|H(p)\rangle  = \frac{1}{2\sqrt{m m'}}\left(
               f_{+}(q^2) (p+p')_{\mu} + f_{-}(q^2) (p-p')_{\mu} \right),
\ee
where $P$ stands for a light pseudoscalar ($\pi $, $K$, $\eta $) and $H$ for
a heavy one ($D$, $B$, $D_{s}$, $B_{s}$),
$m', m$ are the masses of the $P$ and $H$ mesons respectively, $p'$ and $p$ are
their four - momenta. In the
case of the pseudoscalar - vector transitions we have four independent
formfactors \c{lie}:
\be
\la{vpv}
\langle V(\epsilon ,p')|V_{\mu}|H(p)\rangle  = \frac{i}{2\sqrt{m m'}}
                       \frac{g(q^2)}{m+m'}
                       \epsilon_{\mu\nu\rho\sigma }\epsilon^\nu (p+p')^\rho
                                                                (p-p')^\sigma ,
\ee
\begin{eqnarray}
\la{apv}
\langle V(\epsilon ,p')|A_{\mu}|H(p)\rangle  = \frac{1}{2\sqrt{m m'}}
                   \left[ (m+m')
                   f(q^2) \epsilon_{\mu } + \right. \\ \nonumber
     \left.\frac{a_{+}(q^2)}{m+m'} (\epsilon \cdot p) (p+p')_\mu  +
     \frac{a_{-}(q^2)}{m+m'} (\epsilon \cdot p) (p-p')_\mu \right],
\end{eqnarray}
where $V(\epsilon ,p')$ describes the vector particle ($\rho $, $K^{\ast }$,
$\phi $) with momentum $p'$ and polarization $\epsilon_{\mu } $. For vanishing
electron mass the formfactors $a_{-}, f_-$ do not contribute to the decay
probabilities. The formfactors
defined in \r{vpv}, \r{apv} are related to the other
commonly used set (e.g. \c{aba}) by the simple formulae: $g = V$, $f = A_1$,
$a_+ = -A_2$.
The strategy of our calculation  is standard. At first we calculate the
values of the formfactors in the point of maximum four-momentum transfer
$q^2_{max}$ and then
we assume the pole dependence of the formfactor:
\be
\la{fo}
f(q^2) = \frac{f(0)}{1-q^2/m^{\ast 2}},
\ee
where $m^{\ast }$ are: $2.01\, GeV$ for $c \rightarrow q$ ($q = u, d$)
transitions,
$2.11\, GeV$ for $c \rightarrow s$\footnote{We neglect the different masses of
poles in different formfactors in the case of $D$ decay, because the
dependence of the mass $m^{\ast }$ on the decay channel is  known to have
little effect on the results \c{aba}.}. For $B$ decays, in the case of the
$a_+$ and $f$ formfactors we used the pseudovector pole $5.71\, GeV$ \c{cas3}
and for the $g$ formfactor the vector pole $5.32 GeV$ \c{pdg}.

\section{Formfactors at $q^2_{max}$}

In the corrected version of the MIT bag model \c{shur}, \c{iza}
the light mesons are treated conventionally but the heavy ones are built
like the hydrogen atom. The heavy quark defines the center of mass of the
system
and occupies the center of the spherical bag. Considering the limit of the
infinite mass of the heavy quark there is no difference between
the pseudoscalar and the vector heavy mesons,
because the colour magnetic interactions between the components of the
particles
are proportional to the inverse of the heavy quark mass. Similarly, there is
no difference in structure between corresponding mesons with $c$ and $b$
quarks.
E.g. $B$ and
$D$ mesons posses the same structure. These two statements
follow directly from the HQS. The situation is quite different for light
mesons.
In this case there is an
important difference between pseudoscalar and vector particles due to
the colour magnetic interaction of their components. In the language of the
bag model this is expressed by the difference in the bag radius, namely about
$3.4\, GeV^{-1}$ for pseudoscalar mesons and about $4.4 \, GeV^{-1}$ for the
vector mesons \c{iza}. There is also
the difference between particles with the $u$ or $d$ and the $s$ quarks, e.g.
between $K$ and $\rho $ mesons. The last difference is a consequence of the
broken $SU(3)$
flavour symmetry, because the mass of the $u, d$ is approximately 0 and the
mass of the $s$ quark (from fit from \c{iza}) is $0.273\, GeV$. As we shall
show
in the next section, the colour magnetic interaction can explain the
suppression of the ratio \r{R}. The direct influence of the spectator antiquark
explains the similarity between the decay widths of
$D \rightarrow K^{\ast -}e^+\nu_e$ and $D_s \rightarrow \phi e^+\nu_e$.

The matrix elements of the hadronic current in the bag model can be found in
\c{gold}:
\begin{eqnarray}
\la{had}
\langle X(p')|J_{\mu }|H(p)\rangle  = \hskip 8cm\\ \nonumber
\langle X(p')|\int_{Bag} d^3\hbox{{\bf r}}
                  \, e^{-i\hbox{{\bf k}}\cdot \hbox{{\bf r}}}
       \bar{\psi_q}(\hbox{{\bf r}},t) \gamma_{\mu} (1-\gamma_5)
           \psi_Q(\hbox{{\bf r}},t)|H(p)\rangle |_{t=0}\cdot
           \langle \psi_{\bar{q}'}|\psi_{\bar{q}}\rangle |_{t=0}, \hskip-1.1cm
\end{eqnarray}
where $\psi_{q}$ and $\psi_{Q}$ are spinors describing respectively the light
quark and the heavy quark taking part in the weak decay. The bracket at the end
of the formula is the
overlap function of the spectator antiquark in the parent ($\bar{q}$) and
the daughter ($\bar{q}'$) particle. This overlap corresponds to the Isgur-Wise
function in the case of the heavy to heavy meson transitions.
$\hbox{\bf k} = \hbox{\bf p} - \hbox{\bf p}'$ is the momentum transfer in the
process. The integral is over the whole space filled by the (anti)quark fields
at the moment of the decay. In the case of the bag model this is the
intersection
of the bags of the parent and daughter particles. The moment of the decay is
arbitrary and is here chosen to be zero. The integral must
be performed in some well defined reference frame and in principle should
give us the full dependence of the formfactors in terms of the momentum
transfer {\bf k}. Unfortunately, the prescription is not covariant because we
do not know the boost operators acting on the quark fields. Such operators
are necessary, because the spinors $\psi $ are known only in the reference
frame
in which the bag is at rest:
\be
\la{spin}
\psi^0(\hbox{{\bf r}},t) = \left[\begin{array}{c}
                     iF(r)\chi \\
                     G(r) \sigma  \cdot \hbox{\bf r}\chi
                     \end{array} \right].
\ee
Here the functions $F$ and $G$ for the light quark are proportional to the
spherical Bessel functions $j_0$ and $j_1$; $\chi $ is a Pauli spinor.
Generally in the case of the decay it is not possible to find a frame where
both
bags are at rest. For this reason only the calculation for small values of
{\bf k} is reliable. Expanding the right hand side of equation
\r{had} in powers of $|\hbox{{\bf k}}|$ and comparing with \r{pp}, \r{vpv},
\r{apv} one can find the formfactors in the point of maximum four - momentum
transfer $q^2_{max}$ where $\hbox{\bf k} = 0$. The details of these
calculations
can be found in papers \c{lie, hoglie}. As the most useful frame of reference
we
consider the modified Breit frame \c{eeg}. As results we have:
\be
f_+(q^2_{max}) = \frac{m+m'}{2\sqrt{m m'}} [ G_0 - \delta G_2],
\ee
\be
g(q^2_{max}) = \frac{m+m'}{2\sqrt{m m'}}  G_1,
\ee
\be
f(q^2_{max}) = \frac{2\sqrt{m m'}}{m+m'} H_1,
\ee
\be
a_+(q^2_{max}) = \frac{1}{1-\delta^2} \sqrt{\frac{m'}{m}} [-H_0 -
                         (1+ \frac{1}{2}\delta )H_1 - H_2],
\ee
where $\delta = (m - m')/(m + m')$ and
$G_0$, $G_2$, $G_1$, $H_0$, $H_1$, $H_2$ are the Sachs formfactors
defined in \c{lie}. The first two formfactors describe
pseudoscalar - pseudoscalar transitions the next four correspond to
 pseudoscalar - vector transitions. These formfactors are expressed by
the bag integrals:
\be
G_0 = \hat{N} \hat{N}',
\ee
\be
G_2 = -\beta_{Br} \hat{d} \hat{N}',
\ee
\be
G_1 = (\hat{g}_A + \beta_{Br}\hat{\mu}) \hat{N}',
\ee
\be
H_0 = -\beta_{Br} \hat{d} \hat{N}',
\ee
\be
H_1 = \hat{g}_A \hat{N}',
\ee
\be
H_2 = (-\frac{1}{2}\hat{g}_A - \beta_{Br}\hat{\mu } -
       \frac{2}{15}\beta_{Br}^2\hat{B}) \hat{N}',
\ee
where:
\be
\la{n}
\hat{N} = \int d^3\hbox{{\bf r}}[F^{\ast }_q F_Q + G^{\ast }_q G_Q],
\ee
\be
\hat{g}_A = \int d^3\hbox{\bf r}[F^{\ast }_q F_Q - \frac{1}{3}G^{\ast }_q G_Q],
\ee
\be
\hat{\mu } = -\frac{1}{3}(m+m')\int d^3\hbox{\bf r} r [F^{\ast }_q G_Q +
                                                  G^{\ast }_q F_Q],
\ee
\be
\hat{d} = \frac{1}{3}(m+m')\int d^3\hbox{\bf r} r [F^{\ast }_q G_Q -
                                              G^{\ast }_q F_Q],
\ee
\be
\hat{B} = (m+m')^2 \int d^3\hbox{\bf r} r^2 G^{\ast }_q G_Q,
\ee
\be
\la{np}
\hat{N}' = \int d^3\hbox{\bf r}[F^{\ast }_{\bar{q}'} F_{\bar{q}} +
           G^{\ast }_{\bar{q}'} G_{\bar{q}}].
\ee
The last integral describes the overlap of the spectator antiquarks
in the process. $F_q$ and $G_q$ are the upper and the lower component
of spinor \r{spin} describing the light quark produced in the process.
$F_Q$, $G_Q$ describe the heavy quark.
$\beta $ is the so called retardation factor and comes from the relativistic
corrections to the transformation of time:
\be
\beta_{Br} = 1-\frac{E_q+E_Q}{m+m'},
\ee
where $E_q$ and $E_Q$ are the quark eigenenergies in the rest frames of the
bags. Unfortunately it is impossible to calculate the integrals, because
the wavefunction $\psi_Q$ of the heavy quark is not known
within the bag model. Nevertheless, we can still obtain some interesting
results. First of all let us take the infinite mass limit in the bag integrals.
Using the fact that the lower component of the spinor $\psi_Q$
must vanish and parameter $\beta_{Br}$ is very small (because $E_Q \approx m_Q
\approx m >> m'$) one finds that the Sachs formfactors $G_2$, $H_0$ are
negligible and
$G_1 = H_1 = -2 H_2$. In terms of the invariant formfactors this means that:
\be
\la{f+}
f_+(q^2_{max}) = \frac{m+m'}{2\sqrt{m m'}} {\cal N}
\ee
for pseudoscalar - pseudoscalar transitions and
\be
\la{a}
- a_+(q^2_{max}) = g(q^2_{max}) = \frac{m+m'}{2\sqrt{m m'}} {\cal N},
\ee
\be
\la{f}
f(q^2_{max}) = \frac{2\sqrt{m m'}}{m+m'} {\cal N},
\ee
for pseudoscalar - vector transitions. Here ${\cal N}$ is a product of two
overlaps:
first ($\hat{N}$) between the heavy quark and the produced light quark and
second ($\hat{N}'$) between the spectators of the process:
\be
\la{Nn}
{\cal N} = \hat{N}\hat{N}' = \langle \psi_q|\psi_Q\rangle  \cdot
\langle \psi_{\bar{q}'}|\psi_{\bar{q}}\rangle .
\ee
Thus the calculation of all the formfactors is reduced to the
calculation of the quantity ${\cal N}$.  First of all it is straightforward
to see that in the case of the heavy to heavy meson
transition one can reproduce the HQS prediction connected with the Isgur-Wise
function. In this case both overlaps are equal to unity. This is true because
for the decay in the limit $\hbox{\bf k} = 0$ a heavy quark at rest is replaced
by another heavy quark at rest so that $\hat{N} = 1$.
Also $\hat{N}' = 1$, because
the wavefunction of the light antiquark in the meson does not depend
on the heavy quark flavour. In the case of the heavy to light transition,
the overlap of the spectator antiquark
can be calculated easily, but the second overlap requires more attention.
For calculating this overlap let us assume that the decay process proceeds
in two steps:
\begin{itemize}
\item rapid transition between the heavy and the light quark
\item transition of the light quark into its final state
\end{itemize}
The first step is very difficult to calculate. We know neither the heavy
quark wavefunction, nor the light quark, besides the fact that both
wavefunctions are similar. We propose to describe this transition
by the unknown dimensionless function $\alpha (\mu a)$ of the product of
some parameter $\mu $ (dimension of energy) and the radius $a$ of the region
previously occupied by the heavy quark. The radius $a$ should scales like
$1/m_Q$. The parameter
$\mu $ can be constructed only from quantities describing the light quark just
after the transition such as momentum, energy or mass. The second step gives us
more information. Although
we do not know the wavefunction of the heavy quark (or of the light quark just
after transition) we can use the
fact that it occupies the center of the bag. The density of the heavy quark
behaves like a delta function so its wavefunction is the "square root" of
delta.
Let us use for the estimation of the overlap \r{n}
the Gaussian representation of the delta function. Thus the heavy
quark wavefunction is represented by the formula:
\be
\la{hq}
F_Q(r) = Q_{a}(x,y,z) = \frac{1}{(a\sqrt{\pi })^{3/2}} e^{-r^2/2a^2}.
\ee
The overlap of $Q_{a}$ and an arbitrary wavefunction $h(r)$ is:
\be
\la{in}
\int d^3\hbox{\bf r} Q_{a}(\hbox{\bf r}) h(\hbox{\bf r}) \approx 2^{3/2}
             \pi^{3/4}a^{3/2}\, h(0).
\ee
Parameter $a$ has an obvious interpretation as the radius of the region
occupied by the heavy quark.
In the limit $a \rightarrow 0$ ($a$ scales like $1/m_Q$)
$Q_a$ is proportional to a delta function, but integral \r{in} vanishes.
It may seem that there is no interesting information here, but if we want
to calculate the overlap \r{n}, we have to take into account both steps of the
decay. After incorporating \r{in} in \r{n} we find:
\be
\hat{N} = \alpha (\mu a) 2^{3/2} \pi^{3/4} a^{3/2} F^{\ast }_q(0),
\ee
where $F_q$ is the wavefunction of the produced light quark in its final state
in the light meson. Taking the limit $a \rightarrow 0$ and simultaneously
requiring that $\hat{N}$ should be different from zero, we have to assume that:
\be
\alpha (\mu a) = \frac{k}{(\mu a)^{3/2}},
\ee
where $k$ is some constant. Finally:
\be
\la{nov}
{\cal N} = 2^{3/2} \pi^{3/4} k \frac{F^{\ast }_q(0)}{\mu^{3/2}}
\langle \psi_{\bar{q}'}|\psi_{\bar{q}}\rangle ,
\ee
Now one can calculate the quantities, for which the unknown factor in
\r{nov} cancels. Moreover, making
simple assumptions about the parameter $\mu $, one can find the values
of the decay widths.

\section{Numerical results}

\begin{table}[b] 

\caption{Parameters $R^{HV}_{HP}$ (def. (1)).}

\vskip 0.2 cm

\hbox to\hsize{\hss
\begin{tabular}{|l|c|c|c|c|c|c|c|} \hline\hline
 $R^{HV}_{HP}$ & our work & QCD sum  & lattice & chiral & quark    & Exp.   \\
     &          & rules&  \c{aba}& model \c{cas3} & model \c{wir,
bau}&\c{sto}\\
     \hline

$R^{DK^{\ast }}_{DK}$     & 0.56 & $0.5 \pm 0.15 $\c{ball2} &
                          $1.1 \pm 0.6 \pm 0.3$ &  0.56
                          & 1.14 & $0.57 \pm 0.08 $            \\  \hline

$R^{D\rho }_{D\pi }$      & 0.6  & 0.3 \c{ball}     & -     &
                     0.49 & 0.97 &        -          \\  \hline

$R^{D_sK^{\ast }}_{D_sK}$ & 0.71 &        -         &          -           &
                     0.56 &  -   &        -         \\  \hline

$R^{B\rho }_{B\pi }$      & 2.91 & 2.4 \c{ball}  & -  &
                     0.63  & 3.51 &        -          \\

                          & &$1.4 \pm 0.2$ \c{nar} & & & & \\ \hline
$R^{B_sK^{\ast }}_{B_sK}$ & 1.6 &       -          &        -              &
                     0.63 &    -        &        -          \\  \hline\hline
\end{tabular}
\hss}
\end{table}
\begin{table}[b] 
\caption{Polarization parameters \protect $\Gamma_L/\Gamma_T$.}

\vspace*{0.2 cm}

\hbox to\hsize{\hss
\begin{tabular}{|l|c|c|c|c|c|c|c|} \hline\hline
 Decay & our work & QCD sum  & lattice & chiral & quark  & Exp. \\
                     &          & rules&         & model \c{cas3} & Model
                     \c{wir, bau} & \c{sto} \\ \hline

$D \rightarrow K^{\ast } $& 1.03 & $0.86 \pm 0.06$ \c{ball2}
                          & $1.4 \pm 0.3$\c{aba}    &
                            1.31  & 0.89  &  $1.15 \pm 0.17 $\\ \hline

$D \rightarrow \rho  $    & 0.92  & $1.31 \pm 0.11 $ \c{ball}& $1.86 \pm 0.56$
                          \c{lub2}& 1.4    & 0.91  &       -          \\ \hline

$D_s\rightarrow K^{\ast }$& 0.98 &       -          &   -     & 1.25   &  -
&
                                -          \\ \hline

$D_s\rightarrow \phi $    & 1.07 &  -  &$1.49 \pm 0.19$ \c{lub2} & - &  - &
                                -          \\ \hline

$B \rightarrow \rho  $    & 0.78  & $ 0.06 \pm 0.02$ \c{ball}&  -  & 0.36   &
                           1.34  &  -       \\

                          & & $0.125 \pm 0.08$ \c{nar} & & & &  \\ \hline

$B_s\rightarrow K^{\ast }$& 0.43 &      -           &    -    & 0.28   &   -
&
                                -           \\ \hline\hline
\end{tabular}
\hss}
\end{table}

The bag model has some parameters that were adopted from paper \c{iza} in which
they had been
fitted to the spectroscopy of the light and heavy particles. The wavefunctions
of the light quarks are normalized to unity inside the bag. For the light
mesons
we used radii calculated in \c{iza}; in the case of the heavy mesons we take
the
infinite mass limit and we find $R_{D(B)} = 3.9\, GeV$ and
$R_{D_{s}(B_s)} = 4.0\, GeV$. For the decay width calculation we used the pole
form ansatz for the formfactor  dependence on $q^2$ (see equation \r{fo}).

Our knowledge of the formfactors is
up to the unknown factor $k\mu^{-3/2}$, but we have still got the possibility
to calculate quantities, for which this factor cancels. These quantities are:
the polarization factors $\Gamma_L/\Gamma_T$ and the $R^{HV}_{HP}$ parameters
(cf. \r{R}).
The second parameter is calculable only for transitions for which parameter
$\mu $
can cancel. This parameter depends not only on the created light quark but also
can be indirectly influenced by the spectator of the process as we shall show
in a moment. Using formulae \r{f+} - \r{f}, \r{nov} and the assumption of
pole dependence \r{fo}, without any fitting, we obtain the results written in
the Tables I, II.

The results from Tables I, II agree well with the available data and are
comparable with the calculations from other models.
We predict the suppression of the ratios $R^{D\rho }_{D\pi }$ and
$R^{D_{(s)}K^{\ast }}_{D_{(s)}K}$ but not of $R^{B \rho }_{B \pi }$.
This tendency is qualitatively confirmed by QCD sum rules calculations \c{ball}
and partially agrees with chiral model results.
We also predict a large longitudinal polarization in the decays
$B \rightarrow \rho $ and $B_s \rightarrow K^{\ast } $, although not as
large as according to QCD sum rules and chiral model calculations. The lattice
results have large errors and it is difficult to draw from them  definite
conclusions.
\begin{table}[b] 
\caption{Decay widths in units $|V_{cq}|^2 10^{11} s^{-1}$ (upper entries) and
the corresponding branching ratios (lower entries) calculated assuming
$|V_{cd}| = 0.221$, $|V_{cs}| = 0.974$ and life times from \protect\c{pdg}.}
\vspace*{0.2 cm}
\hbox to\hsize{\hss
\begin{tabular}{|l|c|c|c|c|c|c|} \hline\hline
$$Decay$ $ & our work & QCD sum  & lattice & chiral         & quark     & Exp.
\\
$ $    &          & rules&         & model \c{cas3}$^a$
& model \c{wir, bau} & Br. \c{pdg} \\ \hline
$D^0 \rightarrow K^-$        & 0.77  &  \c{ball2}
                                      & $0.54 \pm 0.3 \pm 0.14$& 0.69$^b$
                                      & 0.83     & $(3.31 \pm 0.29)\%$  \\
                                      & 3.24\%&$2.7 \pm 0.6$\%  & \c{aba} &   &
                                      &                      \\ \hline
$D^0 \rightarrow K^{\ast -}$ & 0.43  & $0.39 \pm 0.15$ & $0.64 \pm 0.28$ &
                             0.37  & 0.95     & $(1.7 \pm 0.6)\%  $   \\
              & 1.71\%& \c{ball2}$^{c, d}$  & \c{aba}$^c$ & 1.6\%& & \\ \hline
$D^0 \rightarrow \pi^-$      & 1.9   &$0.8 \pm 0.17$
                             & $1.01 \pm 0.61 \pm 0.2 $ & 1.8
                             & 1.41  &$(0.39^{+0.23}_{-0.12})\% $\\
                             & 0.39\%& \c{ball}  &  \c{aba}       & 0.39\%&
                             &                     \\ \hline
$D^0 \rightarrow \rho^-$     & 1.14  &$0.24 \pm 0.07$
                             & $1.2 \pm 0.61 \pm 0.2$    & 0.92
                             & 1.38  &  -    \\
                             & 0.23\%& \c{ball}    &  \c{aba}       & 0.19\%&
                             &                      \\ \hline
$D_s^+ \rightarrow \Phi^0$   & 0.38  &    -   & $0.44 \pm 0.06$         &
                        -    &  -    &$(1.6 \pm 0.7)\% $\\
                             & 1.6\% &          &   \c{lub2}$^a$            &
                             &               &       \\ \hline
$D_s^+ \rightarrow K^0$      & 1.26  &   -      &      -     & 1.39
                             &  -     &    -              \\
                             & 0.28\%&          &            & 0.31\%
                             &                  &    \\ \hline
$D_s^+ \rightarrow K^{\ast 0}$&0.9   &   -      &     -      & 0.78
                                      &  -    &    -             \\
                                      & 0.2\% &          &           & 0.17\%
                                      &                 &    \\ \hline\hline
\end{tabular}
\hss}
\hskip-1.8cm$^a$ We used $|V_{cs}| = 0.975$ and $|V_{cd}| = 0.222$ for
undoing $\Gamma $ dependence of CKM matrix

\hskip-1.5cmelements.
$^b$ This is $D^- \rightarrow K^0$ channel.
$^c$ This is $D^+ \rightarrow \bar{K^0}^{\ast }$.
$^d$ To get decay

\hskip-1.5cmwidth we used $\tau_{D^+} = 1.07 ps$
\end{table}

For the decay width calculations we need the knowledge of the $\mu $ parameter.
The only thing we know is that it should be described by some observables
connected with the light quark just after the transition. We choose this
parameter with two assumption: to get as good agreement with the data as we can
and, if possible, to remove the constant $k$ (it means to put $k = 1$).
Fortunately we find such a quantity: $\mu = \bar{p} + m_q $, where
$\bar{p}$ is the avarage momentum of the produced light quark just after
the transition and $m_q$ is its mass. Within the bag model the momentum
$\bar{p}$ is calculable as follows. In the point of zero
momentum transfer it has to be equal to the momentum of the heavy quark before
decay. On the other hand in the reference frame chosen, the
momentum of the heavy quark is equal to the momentum of the spectator
antiquark $\sqrt{<\hbox{ \bf p}^2>}$ in the heavy meson. In this way, besides
the overlap of the spectator antiquark
$\hat{N}'$ \r{np}, we have also the additional influence in \r{nov} of the
spectator through the $\mu $ parameter.
This phenomenon causes the decay $D_s \rightarrow \phi $
to be similar to the decay $D \rightarrow K^{\ast }$. With the same
$\mu $ for both processes,
$D_s$ to $\phi $ decays would be about 1.7 times more probable then the
$D$ to $K^{\ast }$ decays.
This is also the case for other "twin" decays as: $D^0 \rightarrow \rho^-$ vs.
$D^+_s \rightarrow K^{\ast 0}$ or $B^0 \rightarrow \pi^+$ vs.
$B^0_s \rightarrow K^+$.
The decay widths are collected in Tables III, IV.
\begin{table}[b] 
\caption{ Decay widths in units $|V_{ub}| 10^{13} s^{-1}$
and corresponding branching ratio calculated assuming
$\tau_B = 1.49 ps$ \protect\c{tau} and $|V_{ub}| = 0.0043$ \protect\c{bucc}.}

\vspace*{0.2 cm}
\hbox to\hsize{\hss
\begin{tabular}{|l|c|c|c|c|c|c|} \hline\hline
Decay  & our work & QCD sum           & lattice & chiral  & quark        &Exp.
\\
     &   & rules &  \c{aba}& model \c{cas3}  & model \c{wir, bau}& \\ \hline

$\bar{B_{}}^0 \rightarrow \pi^+$  & 0.74 &$ 0.51 \pm 0.11$ \c{ball}
                   & $0.9 \pm 0.6$ & 5.43$^c$
                   & 0.74 & $<$ 0.014\%\,\,\,@90\%\,CL \,\c{cle}$^a$  \\
                   & 0.02\%&$0.36 \pm 0.06$ \c{nar} &  & 0.14\%  &   & \\
\hline
$\bar{B_{}}^0 \rightarrow \rho^+$ & 2.18 &$ 1.2 \pm 0.4 $ \c{ball}
                  & $1.4 \pm 1.2$
                  & 3.4 & 2.6  &$(0.103 \pm 0.036 \pm 0.025)\%$ \c{dr}$^b$ \\
                  & 0.06\%&$0.50 \pm 0.22$ \c{nar}&  & 0.087\% &  &   \\
\hline
$\bar{B_s}^0 \rightarrow K^+$     & 0.89 & - &  -         & 5.43
                                     & -     &   -    \\
                                     &0.024\% & & &0.14\% & & \\ \hline
$\bar{B_s}^0 \rightarrow K^{\ast +}$&1.43&      -          &  -   & 3.4
                                     &  -    &   -    \\
                                     &0.04\% & & &0.087\% & & \\ \hline\hline
\end{tabular}
\hss}
\hskip-2.7cm $^a$ This is upper limit on $Br(B \rightarrow \pi l\nu)$.
$^b$ This is $Br(B \rightarrow \rho^0l^-\bar{\nu}_l)$.
$^c$ This is $B^0 \rightarrow \pi^-$ channel.
\end{table}
\begin{table}[b] 
\caption{Formfactors $f_+(0)$ for pseudoscalar - pseudoscalar transitions
at \hbox{$q^2 = 0$.}}

\vspace*{0.2 cm}
\hbox to\hsize{\hss
\begin{tabular}{|l|c|c|c|c|c|c|} \hline\hline
$f_+(0)$  & our work &  QCD sum      & lattice & chiral          & quark
&Exp. \\
          &  & rules     &  & model \c{cas3}  & model \c{wir, bau}& \\ \hline

$f^{DK}_+$    & 0.71 &$0.6^{+0.15}_{-0.1}$ \c{ball2}
              &$0.6 \pm 0.15 \pm 0.07$  \c{aba}
              & 0.67 & 0.76 & $0.76 \pm 0.02$  \c{hai, wit}    \\ \hline
$f^{D\pi }_+$ & 0.8  &$0.5 \pm 0.1$ \c{ball} & $0.58 \pm 0.09$ \c{lub2} & 0.79
              & 0.69 &$0.8^{+0.21}_{-0.14} \c{pdg}$ \\
              & & &$0.84 \pm 0.12 \pm 0.35$ \c{bern} & & & \\ \hline
$f^{D_sK}_{+}$& 0.74 & - & $0.84 \pm 0.08 \pm 0.18$ \c{bern}   & 0.78 & 0.64
              &  -  \\ \hline
$f^{B\pi }_+$ & 0.33 &$0.26 \pm 0.02$ \c{ball}&$0.3 \pm 0.14 \pm 0.05$\c{aba}
              & 0.89 & 0.33 & -  \\
              &      & $0.23 \pm 0.02$\c{nar} & & &  &    \\ \hline
$f^{B_sK}_{+}$& 0.36 &     -         &  -  & 0.89     &  -  & - \\ \hline\hline
\end{tabular}
\hss}
\end{table}

The agreement with the available data is quite impressive. In the case of $D$
decays our decay
widths are almost the same as, those in the chiral model and
agree with the quark model in size (except for the $D \rightarrow K^{\ast }$
decay). The QCD sum rules give smaller
widths for $D \rightarrow \pi $ and $D \rightarrow \rho $ decays. On the other
hand
in the case of $B$ decays our results are rather closer to the QCD sum rules
predictions (for $B \rightarrow \pi $ almost within errors \c{ball}) and do not
agree with the chiral model calculations which give much larger widths.
One can also compare the values of the formfactors at $q^2 = 0$.
These numbers are collected in Tables V, VI.
\begin{table}[b] 
\caption{Formfactors for pseudoscalar--vector transitions at $q^2 = 0$.}
\vspace*{0.2 cm}
\hbox to\hsize{\hss
\begin{tabular}{|l|c|c|c|c|c|c|c|} \hline\hline
form -   & our work & QCD sum   & lattice & chiral          & quark    &Exp. \\
factors  &          & rules     &       & model \c{cas3}  & model \c{wir, bau}&
 \c{cas3, kod, anj} \\ \hline\hline
\multicolumn{7}{|c|}{$D \rightarrow K^{\ast }$}    \\ \hline\hline
f(0)      & 0.55 &$0.5 \pm 0.15$\c{ball2}&$0.64 \pm 0.16$\c{aba}& 0.48 & 0.88 &
$0.48 \pm 0.05$     \\ \hline
$- a_+(0)$& 0.63 &$0.6 \pm 0.15$&$0.4 \pm 0.28 \pm 0.04$  & 0.27  & 1.15
&$0.27 \pm 0.11$ \\ \hline
g(0)      & 0.63 &$1.1 \pm 0.25$&$0.86 \pm 0.24$         & 0.95   & 1.23
&$0.95 \pm 0.2$  \\ \hline\hline
\multicolumn{7}{|c|}{$D \rightarrow \rho  $}        \\  \hline\hline
f(0)      & 0.69 &$0.5 \pm 0.2$ \c{ball}&$0.45 \pm 0.04$\c{lub2}
& 0.55 & 0.78 & - \\ \hline
$-a_+(0)$ & 0.83 &$0.4 \pm 0.1$&$0.02 \pm 0.26$  & 0.28 & 0.92 & - \\ \hline
g(0)    & 0.83 &$1.0 \pm 0.2$&$0.78 \pm 0.12$  & 1.01 & 1.23 & - \\
\hline\hline
\multicolumn{7}{|c|}{$D_s \rightarrow K^{\ast }$}  \\  \hline\hline
f(0)      & 0.61 &  -  &  -  & 0.52 & 0.717  &   -  \\ \hline
$-a_+(0)$ & 0.71 &  -  &  -  & 0.3  & 0.853  &   -  \\ \hline
g(0)      & 0.71 &  -  &  -  & 1.08 & 1.250  &   -  \\ \hline\hline
\multicolumn{7}{|c|}{$D_s \rightarrow \phi $}  \\ \hline\hline
f(0)      & 0.53 &  -  & $0.52 \pm 0.03$\c{lub2}  &  -   & 0.820  & - \\ \hline
$-a_+(0)$ & 0.59 &  -  & $0.17 \pm 0.17$  &  -   & 1.076   & - \\ \hline
g(0)      & 0.59 &  -  & $0.86 \pm 0.1$   &  -   & 1.319  &   -  \\
\hline\hline
\multicolumn{7}{|c|}{$B \rightarrow \rho   $}       \\  \hline\hline
f(0)      & 0.27 &$0.5 \pm 0.1$ \c{ball}&$0.22 \pm 0.05$\c{aba}  & 0.21 & 0.28
&  -  \\
      & &  $0.38 \pm 0.04$\c{nar}  &  & & &  \\ \hline
$-a_+(0)$ & 0.61 &$0.4 \pm 0.2$         &$0.49 \pm 0.21 \pm 0.05$    & 0.2
& 0.28 &  -  \\
      & &  $0.45 \pm 0.05$  &  & & &  \\ \hline
g(0)      & 0.46 &$0.6 \pm 0.2$         &$0.37 \pm 0.11$ & 1.04 & 0.33 &  -
 \\
     & &  $0.45 \pm 0.05$  &  & & &  \\ \hline\hline
\multicolumn{7}{|c|}{$B_s \rightarrow K^{\ast }$}  \\  \hline\hline
f(0)      & 0.25 &  -  &  -  & 0.2    & 0.328  &  -  \\ \hline
$-a_+(0)$ & 0.5  &  -  &  -  & 0.21    & 0.331  &  -  \\ \hline
g(0)      & 0.38 &  -  &  -  & 1.08    & 0.369  &  -  \\ \hline\hline
\end{tabular}
\hss}\vskip-0.65cm
\end{table}

In the case of $f_+$ formfactors we agree with the QCD sum rules calculations,
except for the $D \rightarrow \pi $ transition where our result is larger.
We also have good agreement with
the quark model for all calculated decays and with the chiral model predictions
for $D$ decays. For $B$ decays the chiral model gives larger values than any
other model considered here.
In the case of the formfactors for the pseudoscalar - vector transitions,
the situation is more
complicated. Our results for the $f$ formfactors roughly agree with other
models
and only in $B$ decays the QCD sum rules calculations
give a little bit larger numbers. In the case of the $a_+$ and $g$ formfactors
we find that in the infinite mass limit they are equal at the point of maximum
four - momentum transfer. From this fact and from the pole dominance assumption
it follows that
they are also similar at zero four - momentum transfer. This is not the case
for other models (except of the quark model). If the experimental results and
other model calculations do
not change, this would mean that finite heavy quark mass corrections are
quite large for the $a_+$
and $g$ formfactors. Moreover, comparing the numbers one can risk the
conclusion
that mass corrections decrease the values of
$- a_+$ and increase the values of the $g$. For $B$ decays corrections
should be smaller. From comparison with the available data it follows that
the corrections to the decay widths are little affected by these changes.
One may hope that this may be also true for other not yet measured processes.

We also checked our predictions for another choice of the parameter $\mu $
putting it equal to the
energy of the light quark $\mu = \sqrt{\bar{p}^2 + m_q^2}$.
We find $k = 0.66$ from fitting to the $D \rightarrow K$ decay.
The only difference in the results for this new choice of $\mu $,
as compared to the previous one, was the suppression of $Q \rightarrow q$
transitions ($Q = b, c$ and
$q = u, d$) relative to $Q \rightarrow s$ transitions. The formfactors
describing
$Q \rightarrow q$ transitions are smaller by a factor $k = 0.66$.
It means that our prediction for $D \rightarrow K^{\ast }$ and
$D_s \rightarrow \phi $
do not change. For the $D \rightarrow \pi $ decay we found
$f_+(0) = 0.53$  and for the
$B \rightarrow \pi $ decay $f_+(0) = 0.22$. These predictions are quite similar
to the QCD sum rules results \c{ball}, \c{nar}, however, our first choice
agrees
better with the $D \rightarrow \pi $ decay data.

In conclusion, we performed an analysis of the semileptonic decays of heavy to
light mesons in the framework of the corrected MIT bag model. We explain the
suppression
of pseudoscalar--vector transition $D \rightarrow K^{\ast }$ relative to the
pseudoscalar--pseudoscalar transition $D \rightarrow K$ as an effect of the
colour magnetic interaction between
the produced light quark and the antiquark spectator.
We can also identify the source of the restored $SU(3)$ flavour
symmetry in the case of the "twin" decays e.g.
$D_s \rightarrow \phi $ and $D \rightarrow K^{\ast }$ as an additional
influence
of the spectator antiquark in the decay "two step" process and  of momentum
conservation. Additionally, we calculated the decay widths of several
semileptonic processes for heavy--light transitions. Agreement with the
experimental data, whenever available is good.

\vspace{0.4 cm}
{\bf Acknowledgements}

\vspace{0.2 cm}
I wish to thank Professor K. Zalewski and
Professor H. H\o gaasen for helpful discussions. I am also grateful to
Professor K. Zalewski for his careful reading of the manuscript.

\end{document}